# Operation of PAL-XFEL LLRF

J. Hu[†], J. Hong, K.H. Kim, S. Jung, D. Na, Y. J. Park, S.-H. Kim, S. Park, C.-K. Min, H.-S. Kang, H.-S. Lee, PAL, POSTECH, Pohang, South Korea

*Abstract*

PAL-XFEL (Pohang Accelerator Laboratory X-ray Free Electron Laser) started RF conditioning in October 2015 and has been operating reliably for ~ 4 years. The machine's LLRF and SSA systems contributed to the stable operation of PAL-XFEL with over 99% availability. The LLRF and SSA systems showed some problems in rare cases. The delay caused by the problem is very small, but PAL-XFEL can stop working. Some issues have been identified and resolved. We want to share the experience.

## INTRODUCTION

PAL-XFEL started its operation by RF conditioning in October 2015 and has been operating reliably for about 4 years. PAL-XFEL layout is shown in Figure 1 and its main parameters and linac parameters are shown in Table 1 and 2. As shown in the tables, PAL-XFEL is a hard/soft X-ray FEL machine based on S-band normal-conducting linac. The linac comprises with 51 RF stations, each of which has 1~4 cavities for acceleration of electron beam. Each RF station is composed of 1 klystron, 1 high-voltage (HV) modulator, 1 SSA (Solid-State Amplifier), 1 LLRF (Low-Level RF) and 1 temperature-controlled rack (TC Rack).

Table 1: Main parameters. Updated from ref. [1].

| Parameter | Unit | Design (Hard/Soft) | Hard/Soft |
|---|---|---|---|
| Charge/bunch | nC | 0.2 | 0.18 |
| Slice emittance | mm·mrad | 0.5 | 0.55 |
| e- energy | GeV | 10/3.15 | 9.47/3.0 |
| Pulse repetition | Hz | 60 | 60 |
| e- peak current | kA | 3.0/2.5 | 2.5/2.2 |
| FEL gain length | m | 3.6/1.8 | 3.61/2.08 |
| FEL λ | nm | 0.1/1.0 | 0.104/1.52 |
| Photons/pulse | $10^{11}$ | 1.8 | 10.0 |

___________________
* Work supported by MSIT, South Korea
† hjy@postech.ac.kr

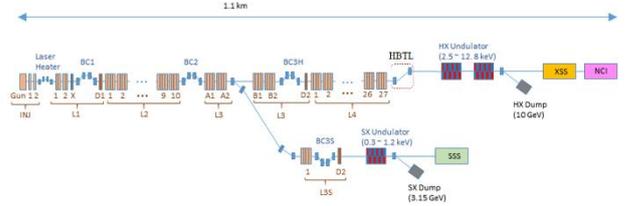

Figure 1: PAL-XFEL layout.

Table 2 : Linac parameters.

| Parameters | Value |
|---|---|
| Type | Normal-conducting |
| Frequency | S-band (2.856 GHz) X-band (11.424 GHz, linearizer) |
| Gun | S-band 1.6 cell photo-gun |
| ACC | S-band 2/3 π mode |
| RF station # | 51 ( 50 : S-band, 1 : X-band) |
| Cavities/klystron | 1 ~ 4 |
| Form of RF station | 1 klystron(25~80 MW, ≤ 4 us), 1 High-Voltage(HV) modulator, 1 SSA(Solid State Amplifier), 1 LLRF & 1 temperature-controlled rack (TC Rack) |

Figure 2 shows the energy of the electron beam at the linac end (HBTL), which is supplied for user experiment this year. From the figure, the electron beam are regulated well to have the energy jitter of about RMS 0.01%.

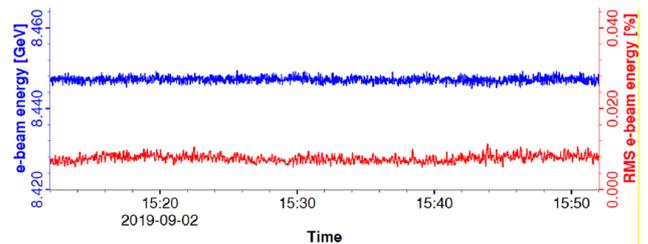

Figure 2 : The electron-beam energy at the HBTL this year.

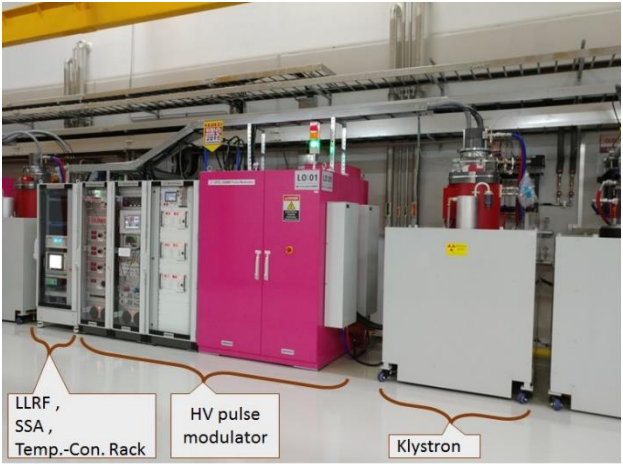

Figure 3 : Picture of a RF station

Figure 3 shows a picture of a RF station of PAL-XFEL. A klystron, a HV modulator and a TC rack are indicated. Within the rack, a LLRF unit and a SSA unit are installed. In this paper, we would like to present the operation status and the related issues of PAL-XFEL LLRF, SSA and TC rack systems.

## AVAILABILITY

Quality of a machine can be evaluated by performance and reliability. PAL-XFEL LLRF performance is presented at the last LLRF Workshop 2017[2]. As a measure of reliability, availability is defined as follows.

A "full system" is considered as single body consisted of 51 units corresponding to each RF station. So, a defect of a certain unit is counted as a defect of the full system. In this way, a station availability and a system availability are considered as follow.

T_tot_plan ≡ the total time planned

(Total period − Maintenance period)

T_tot_def_st ≡ the total delayed time due to defects of a specific unit.

(Station) Availability(%) ≡ $(1 - \frac{T\_tot\_def\_st}{T\_tot\_plan}) * 100(\%)$

(System) Availability(%) ≡
$(1 - \frac{\sum_{st.=1}^{51}(T\_tot\_def\_st)}{T\_tot\_plan}) * 100(\%)$

∴ (System) Availability ≤ (Station) Availability

System availability is not equal to beam availability because the PAL-XFEL beam can usually be supplied with normal stations when some stations are down (about 5~10 stations are spares).

With the definition of availability, the availability of PAL-XFEL LLRF, SSA, and TC rack is shown in Table 3 and Figure 4. From Table 3, the system availability of LLRF and SSA exceeds 99% this year and that for TC rack is 97.8% this year. From Figure 4, there have been no problems except some stations this year.

Table 3: System availability

| System | (System) Availability for 2019 | |
|---|---|---|
| LLRF | 99.6 % | 99.4 % / 97.2 % |
| SSA | 99.8% | |
| TC Rack | 97.8 % | |

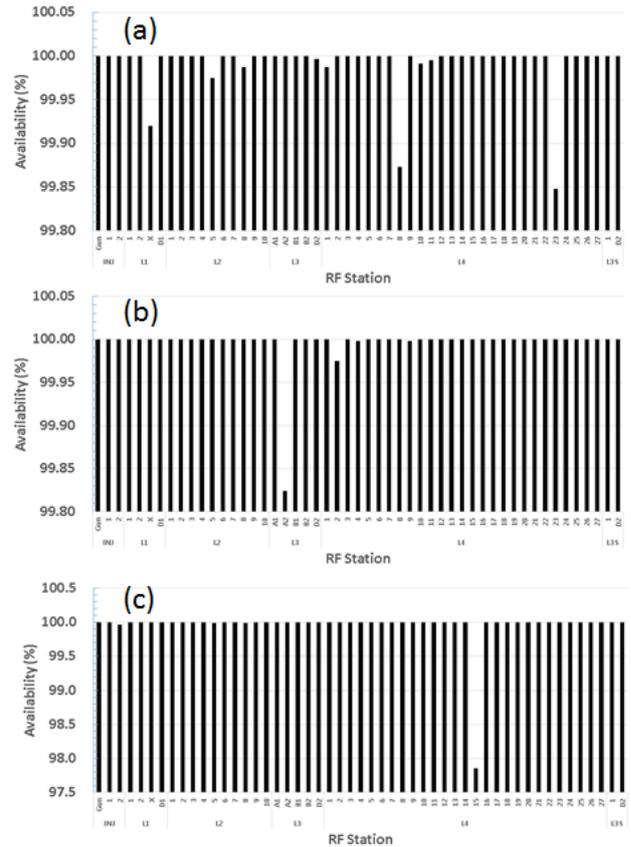

Figure 4: Station availability of LLRF (a), SSA (b) and TC rack (c)

## OPERATION ISSUES

As shown in the previous section, the availability of PAL-XFEL LLRF, SSA and TC rack is highly acceptable as a user-service machine. Though delay time due to the defects of those systems is very small, the defects can stop operation of the machine. So, attempts were devoted to solve the related issues and some problems are identified and solved.

### Issue #1(LLRF) : Freezing of PPC

Figure 5 shows the block diagram of PAL-XFEL LLRF [2]. In the figure, PSI means Personal Safety Interlock system, MIS means Machine Interlock System, EVR means Event Receiver, RF-DS or RFDS means Reference Distribution System, CA means Ethernet for EPICS[3] Channel

Access, FB or F/B means Ethernet focussed on Feedback and Industrial PC or panel PC is abbreviated by PPC and power supply is abbreviated by PS. In PPCs of PAL-XFEL LLRF system, there were some problems as follow.

The symptoms were disconnection of EPICS channel access, PPC freezing and PPC shutdown. The frequency of occurrence was about once in a few months and the occurred stations were random. The first aid applied was rebooting of the problematic PPCs, which usually recovered the PPCs. In case of irrecoverable LLRF sets the replacement of the LLRF sets with a spare sets.

With some investigation it was identified that the problems came from the degradation of power supplies of the PPCs. To see why the PSs were degraded, a simple current analysis was done as shown in Figure 6. It was judged from the analysis that the occasional use of the power supplies higher than the recommended value (3.5A, 70% of the indicated specification) caused the reduction of the lifetime of the power supplies.

All the PPCs' PSs were upgraded Feb. 2019. After the upgrade, the problematic sets were recovered but some sets which had not shown the problems newly showed the symptoms (Table 4). Therefore, we will be conducting further investigation.

Table 4: Occurrence frequency of PPC freezing before/after the PPC-PS upgrade

| RF station | Before | After |
|---|---|---|
| L1-1 | Irrecoverable | |
| XLIN | 1 | 1 |
| L2-1 | 1 | |
| L2-3 | Several times | |
| L2-5 | | 1 |
| L2-8 | | 1 |
| L2-10 | 1 | |
| L4-1 | | 1 |
| L4-10 | | 1 |
| L4-15 | 1 | |
| L4-20 | Irrecoverable | |
| L4-23 | Irrecoverable | |

*Issue #2(LLRF) : Incomplete connection*

PAL-XFEL LLRF had symptoms such as no output power, PSK (Phase Shift Keying) action working incorrectly, and incorrect temperature display. They had common features like normal operation of LLRF until the moment of failure and random occurrence. In total, there were less than 5 times for 4 years. Replacements of LLRF sets were made whenever the incidents occurred.

Symptoms that seemed to be irrelevant were due to incomplete connection of connectors. Figure 7 shows the internal picture of a PAL-XFEL LLRF unit. As one can see from the figure, many RJ45 connectors and 2 HDMI connectors are used.

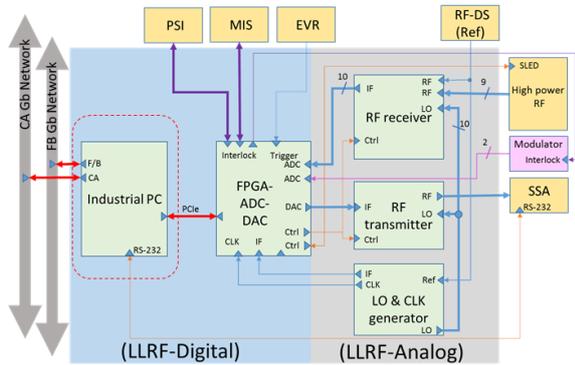

Figure 5: LLRF structure. PPC is emphasized by dotted line.

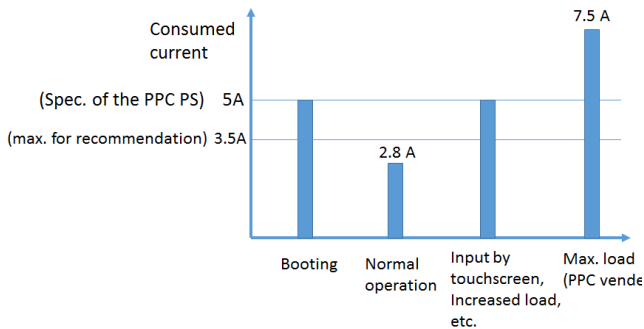

Figure 6: A simple current analysis of a PPC of a PAL-XFEL LLRF set.

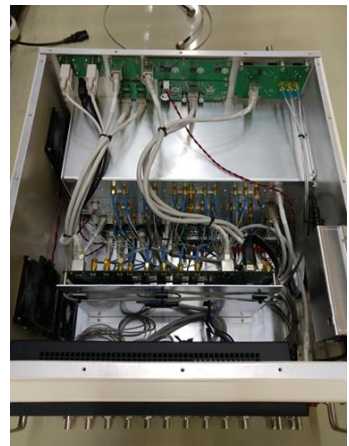

Figure 7: Internal view of a PAL-XFEL LLRF set.

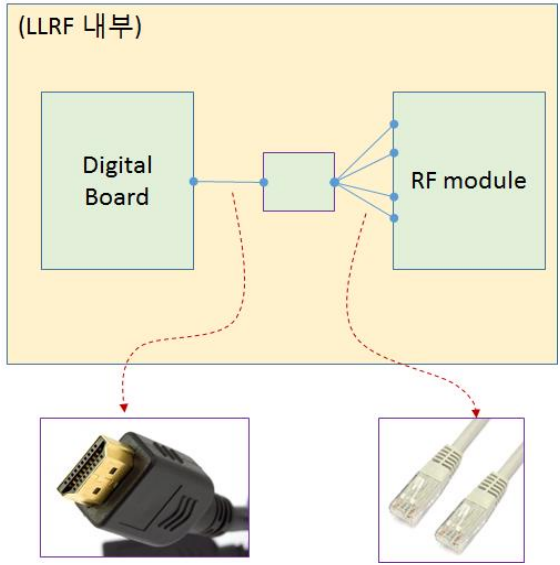

Figure 8: Block diagram illustrating the connections between the digital board and the RF module of PAL-XFEL LLRF.

Figure 8 shows a block diagram of the connection of cables between a digital board and a RF module illustratively. This configuration enables the internal connection simple, very efficient and high-speed operation but not perfect. Through reconnection of the cables, all the sets were recovered. We will search for better solutions such as new venders or new types of connections.

### Issue #3(SSA) : Abnormal Interlock

There are 10 interlock items in PAL-XFEL SSA (Solid State Amplifier or Preamplifier). Occasionally, some interlocks occurred such as "over repetition" or "over current". Sometimes RS-232 communications are lost between LLRF and SSA pairs. The frequency of the events was once in 6~12 months. Usual first aid is as follows. When a SSA interlock occurs, operators review the event and release the interlock in PAL-XFEL control room. When RS-232 communication is lost, the SSA is rebooted. The causes of the events were not clear.

This year important hints were found as follow. At the L3-A2 SSA, "over current" interlock occurred several times and was replaced with a spare set in January 2019. The SSA that was taken out was tested in a test room. There was no interlock when tested in the room. Therefore it was considered that the interlock occurred in the klystron gallery was a fake interlock due to electrical noise.

Another hint was found at L4-2 SSA, which occurred "over repetition" interlock several times suddenly right after winter maintenance. The job in the period was reinstallation of LLRF and reconnection of cables at the site. Therefore, the cable rerouting as shown in Figure 9 has been applied and no problem has occurred so far.

It is inferred from these hints that the strong electrical noise from the HV modulator affected RS-232 communication. For L4-2 SSA case, the abnormal interlock was disappeared in the change as shown in Figure 9 but for some other stations, the reverse showed positive effects. As long-lasting solutions, noise countermeasures will be in place or more noise-resistant communications will be applied in the next version.

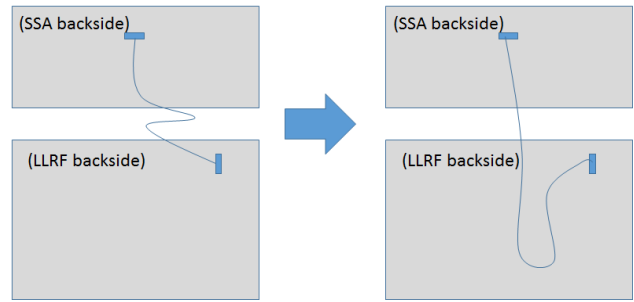

Figure 9 : Rerouting cable between LLRF and SSA pair

### Issue #4(TC Rack) : Water Leakage

TC rack (Temperature-Controlled Rack) for PAL-XFEL is shown in Figure 10. The TC rack regulates temperature inside the rack with cooling water and has a leakage detector to cut electricity when leak events occur. The leak detection system worked properly and helped to detect leakage remotely early.

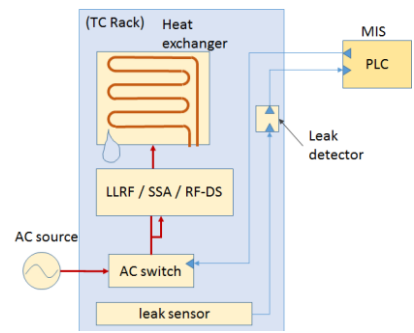

Figure 10: Block diagram of PAL-XFEL TC rack

Table 5: Leakage of TC rack

| Year/Month<br>Station | 2018<br>1~12 | 2019<br>1~8 |
|---|---|---|
| L0-02 |  | 1 |
| L2-02 | 1 |  |
| L2-04 | 1 |  |
| L2-05 |  | 1 |
| L2-08 |  | 1 |
| L4-15 |  | 2 |
| Sum | 2 | 5 |

As shown in Table 5 leak events increased considerably this year. Whenever cooling water leak, the leak exchangers are replaced with spare exchangers. The working time required for each replacement is about 1~2 hours and the number of spare exchangers is only two, which are not enough to manage the events.

Therefore, we proceed in three ways; preparing more spare exchangers, redesigning heat exchangers and developing a rack easier to replace heat exchanger.

## SUMMARY

PAL-XFEL has been operating successfully for 4 years. Its LLRF and SSA systems have worked reliably with system availability more than 99% (97% when included TC rack). 4 operating issues with small percentage but critical are shared.